\documentclass[twocolumn,preprintnumbers,superscriptaddress,endnote,nofootinbib,prl]{revtex4}
\usepackage{graphicx}
\usepackage{amsmath, amssymb}

\newcommand{\vev}[1]{\langle {#1} \rangle}

\newcommand{\gsim}{\gtrsim}

\newcommand{\ord}[1]{\mathcal{O}{(#1)}}
\newcommand{\beq}{\begin{equation}}
\newcommand{\eeq}{\end{equation}}
\newcommand{\bea}{\begin{eqnarray}}
\newcommand{\eea}{\end{eqnarray}}

\newcommand{\eq}[1]{Eq.~(\ref{#1})}

\newcommand{\diphot}{\gamma\gamma}

\begin{document}

\pagestyle{plain}


\title{\boldmath Electroweak Phase Transition, Higgs Diphoton Rate, and New Heavy Fermions}

\author{Hooman Davoudiasl
}
\author{Ian Lewis
}
\affiliation{Department of Physics, Brookhaven National Laboratory,
Upton, NY 11973, USA}

\author{Eduardo Pont\'{o}n
}
\affiliation{Department of Physics, Columbia University, New York,
NY 10027, USA}
\affiliation{Department of Physics, Brookhaven National Laboratory,
Upton, NY 11973, USA}

\begin{abstract}

We show that weak scale vector-like fermions with order one couplings
to the Higgs can lead to a novel mechanism for a strongly first-order
electroweak phase transition (EWPhT), through their tendency to drive
the Higgs quartic coupling negative.  These same fermions could also
enhance the loop-induced branching fraction of the Higgs into two
photons, as suggested by the recent discovery of a $\sim 125$~GeV
Higgs-like state at the CERN Large Hadron Collider (LHC).  Our results
suggest that measurements of the diphoton decay rate of the Higgs and
its self coupling, at the LHC or perhaps at a future lepton collider,
could probe the EWPhT in the early Universe, with significant
implications for the viability of electroweak baryogenesis scenarios.

\end{abstract}
\maketitle

The new boson, recently discovered by the CERN Large Hadron Collider
(LHC) experiments at 125~GeV \cite{ATLASHiggs,CMSHiggs}, has
properties very similar to those of the Standard Model (SM) Higgs
boson associated with electroweak symmetry breaking (EWSB).  While the
current data is inconclusive, the measured properties of the new
particle, henceforth referred to as the Higgs and denoted by $H$, seem
to show some mild deviations from the SM expectations.  In particular,
a hint for the deviation in the loop-induced Higgs diphoton decay
($H\to \gamma \gamma$) rate could be caused by new heavy particles,
likely within the reach of the LHC, that couple to the Higgs with
$\ord{1}$ strength.  Vector-like fermions, {\it i.e.}~endowed with
electroweak preserving masses, are leading candidates \cite{VF,ABDF}
given that their masses may naturally be around the weak scale.  It is
interesting to investigate whether these fermions could play a role in
addressing some of the open questions in the SM.

In this work we point out that the above vector-like fermions could
also lead to a strongly first-order electroweak phase transition
(EWPhT) in the early Universe.  In typical models of electroweak
baryogenesis (EWBG), such a strong phase transition is responsible for
the requisite decoupling of baryon number violating processes in the
broken phase (with $\vev{H}\neq 0$).  These baryogenesis scenarios are
quite interesting, as they could in principle be directly tested in
TeV-scale collider experiments.  Our work hence provides a potential
connection between the rate for $H\to \gamma \gamma$ and early
Universe cosmology.

The possibility of strengthening the EWPhT through heavy fermions
coupled to the Higgs was first considered in Ref.~\cite{CMQW}, where
the authors showed that, contrary to the usual lore, new weak scale
bosons are not necessary for that purpose.  The mechanism proposed in
Ref.~\cite{CMQW} relies on a transfer of entropy from decoupling
fermionic degrees of freedom after EWSB. Here, we point out that
vector-like fermions can lead to a strongly first-order EWPhT through
a combination of a negative quartic coupling (at finite temperature)
together with stabilizing higher-dimension operators.  Such a
mechanism is distinct from those involving new weak scale bosons
coupled to the Higgs that lead to cubic terms (e.g.
Refs.~\cite{Espinosa:1993yi,Carena:1996wj,EWPTDP}).  Instead, the
mechanism underlying the phase transition is closely related to that
proposed in~Ref.~\cite{Grojean:2004xa}, and studied in greater detail
in Ref.~\cite{Delaunay:2007wb}.  The vector-like fermion system, that
may have a rather interesting connection to the Higgs diphoton rate,
can be regarded as a realization of features postulated in
Ref.~\cite{Grojean:2004xa}.  We will come back to the connection
between our setup and other previous works at the end.

To give a simple picture of how the mechanism works, let us first
consider a phenomenological description that captures the dominant
features.  For this purpose, we will ignore the cubic term that can
arise from the SM bosonic degrees of freedom and concentrate on the
following terms in the 1-loop thermal potential $V(\phi, T)$
\bea
V(\phi, T) &\sim& \frac{1}{2} \mu^2 \phi^2 + \frac{1}{4} \lambda \phi^4 + \frac{1}{6} \gamma \phi^6~,
\label{Vtoy}
\eea
for the background field $\phi$ associated with the Higgs.  All of
these terms are temperature-dependent, although we do not indicate it
explicitly.  We will discuss later a specific model that realizes this
scenario with the dimension-6 term being positive.  As we will see,
the crucial feature is that the quartic coupling $\lambda$ can become
negative at finite temperature.  Since at sufficiently high
temperature the mass term becomes positive, we will assume a situation
where $\mu^2>0$, $\lambda < 0$ and $\gamma >0$.  In this case, there
is a local minimum at the origin, separated by a barrier from a second
minimum at $|\phi| \sim \sqrt{-\lambda/\gamma}$.  The associated
potential energy contribution from the $\phi^4$ and $\phi^6$ terms is
of order $\lambda^3/(12\gamma^2)$.  This minimum becomes degenerate
with the minimum at the origin when $\lambda^2 \sim 6 \gamma \mu^2$,
which will determine the critical temperature through the
$T$-dependence of these parameters.  We see that when $\gamma$ is
suppressed by a high mass scale, this occurs when $|\lambda|$ is
small, which can happen for relatively low temperatures.  In addition,
we can estimate the vacuum expectation value (vev) at the critical
temperature as $\phi_c \sim \sqrt{\mu} / \gamma^{1/4}$, which can be
relatively large.  This suggests that, provided the conditions above
can be realized, the ratio $\phi_c / T_c$ can be sizeable, and one can
expect a strongly first-order EWPhT.

By contrast, when the barrier (and therefore the first-order phase
transition) is driven by a cubic term, $E T |\phi|^3$, one has
$\phi_c/T_c \sim E/\lambda$, where $E$ is typically small.  For
instance, in the SM, $E\sim (4 \pi \alpha_W^3)^{1/2}$ is far too small
since $\sqrt{4 \pi \alpha_W}$ is a weak coupling constant.  This is
one of the reasons why EWBG is widely regarded as requiring physics
beyond the SM, so as to enhance the size of $E$.

The desired features can arise as follows.  Consider adding to the SM
a single new vector-like fermion pair $(\chi, \chi^c)$ with a
(vector-like) mass $m_\chi$, that couples to the Higgs field via the
dimension-5 operator
\beq
\Delta {\cal L} = 2 G_m H^\dagger H \, \chi \chi^c + {\rm h.c.}~,
\label{dim5}
\eeq
where $G_m$ is a coefficient with mass dimension -1.  It will prove
useful to present our results from an effective field theory (EFT)
perspective that is, as much as possible, independent of any
particular UV completion, although later on we will give a simple UV
model that leads to Eq.~(\ref{dim5}) with $G_m > 0$.  We have
notationally assumed above that $\chi$ is an $SU(2)_L$ singlet since
such a new fermion will be subject to relatively mild constraints from
EW precision measurements.  However, our formalism will apply with
trivial modifications when $\chi$ is a doublet, with the appropriate
contractions with the Higgs fields in Eq.~(\ref{dim5}).

An immediate consequence of the above operator is that in the presence
of a Higgs vev $\phi$ ($= v =246~{\rm GeV}$ at zero temperature), the
$\chi$ mass becomes
\beq
m_1(\phi) = m_\chi - G_m \phi^2~.  
\label{m1}
\eeq
We will be interested in the 1-loop effective potential for $\phi$,
which will receive a contribution from $\chi$ through $m_1(\phi)$.  In
addition, the interaction in Eq.~(\ref{dim5}) induces divergences in
the Higgs sector corresponding to ``tree-level" operators
\beq
V_0(\phi) = V_{\rm tree} + \frac{1}{6} \bar{\gamma} \phi^6 + \frac{1}{8} \bar{\delta} \phi^8~,
\label{V0}
\eeq
where $\bar{\gamma}$ and $\bar{\delta}$ are free parameters from an
EFT point of view (the bars indicate that we will be thinking of them
as being defined in the $\overline{{\rm MS}}$ scheme).  We have
denoted by $V_{\rm tree}$ the usual quadratic and quartic Higgs terms.
In preparation for the analysis of the high-temperature properties, it
is convenient to impose on the effective potential the conditions
$V'(v) = 0$ and $V''(v) = m^2_H$ at $T=0$, where $v$ and
$m_H$ are the Higgs vev and mass, respectively~\cite{Anderson:1991zb}.
This trades the squared mass parameter and Higgs quartic coupling in
$V_{\rm tree}$ for $v$ and $m_H$, while fixing the 1-loop contribution
due to the new fermion to
\bea
V_1(\phi) &=& - \frac{4}{64\pi^2} \, m^4_1(\phi) \, {\rm Ln}\left( m_1^2(\phi) \right) 
\nonumber \\
&+& \frac{1}{2} \alpha(m^2_1(v)) \, \phi^2 + \frac{1}{4} \beta(m^2_1(v)) \, \phi^4~,
\eea
where ${\rm Ln}(\omega) \equiv \ln \left(\omega/\mu^2 \right) -
\frac{3}{2}$, with $\mu$ the renormalization scale, and
\bea
\alpha(\omega) &=& -\frac{4}{64 \pi^2} \left\{
\left(-3\, \frac{\omega \omega'}{v} + \omega'^2 +
\omega \omega'' \right) {\rm Ln}(\omega)
\right.
\nonumber \\
&-& 
\left.
\frac{3}{2}\left(\frac{\omega \omega'}{v} - \omega'^2 -
\frac{1}{3}\omega \omega''\right) \right\}
+ \bar{\gamma} v^4 + 2 \bar{\delta} v^6~,
\label{alpha}
\\
\beta(\omega) &=& - \frac{4}{128 \pi^2 v^2 } \left\{
2\left(\frac{\omega \omega'}{v} - \omega'^2 -
\omega \omega'' \right) {\rm Ln}(\omega)
\right.
\nonumber \\
&+& 
\left.
\left(\frac{\omega \omega'}{v} -3\, \omega'^2 -\omega \omega''\right) \right\}
-2 \bar{\gamma} v^2 - 3 \bar{\delta} v^4~.
\label{beta}
\eea
This generalizes the expressions derived in Ref.~\cite{CMQW} to the
non-renormalizable case that involves the new parameters
$\bar{\gamma}$ and $\bar{\delta}$, with the explicit factor of $4$
counting the new fermionic degrees of freedom.

Adding the temperature-dependent contributions (for a discussion of
the relevant formalism see for example Ref.~\cite{Quiros:1999jp}), the
free energy reads
\bea
{\cal F} &=& V(\phi) + {\cal F}_1(\phi, T)~,
\label{F}
\eea
where $V(\phi) = V_0(\phi) + V_1(\phi)$ is the zero-temperature
potential discussed above, now including the well-known SM
contributions from the top quark and weak gauge
bosons.\footnote{Although not included in Fig.~\ref{fig1} below, the
$125~{\rm GeV}$ Higgs gives a subdominant effect in the regions of
interest.  We note here, however, that the Higgs slightly strengthens
the EWPhT by lowering $T_c$.  This conclusion is based on the
formalism of Ref.~\cite{Cahill:1993mg}, which uses as a source the
operator $J H^\dagger H$ instead of the usual linear coupling $J H +
{\rm h.c.}$.  This leads to a \textit{real} 1-loop potential
\textit{everywhere}, while maintaining the desired features of the
standard effective potential.  Also, the Nambu-Goldstone bosons from
EWSB do not contribute to the potential.  } The 1-loop thermal
function ${\cal F}_1$ is given by
\beq
{\cal F}_1(\phi, T) = \sum_i \frac{g_i T^4}{2 \pi^2} I_\mp\left(\frac{m_i(\phi)}{T}\right)~,
\label{F1}
\eeq
with $I_\mp(z) = \pm \int_0^\infty \!  dx\, x^2 \ln(1 \mp
e^{-\sqrt{x^2 + z^2}})$.  The index $i$ runs over all the particles,
with number of degrees of freedom given by $g_i$, whose masses
$m_i(\phi)$ depend on $\phi$; the upper and lower signs correspond
to bosons and fermions, respectively.

When the new fermion is at the EW scale, it may be appropriate to use
the high-temperature expansion of Eq.~(\ref{F1}), thus resulting in a
potential, presented in the appendix, that is polynomial in $\phi$
with $T$-dependent coefficients.  However, for the numerical analysis
we will use the full 1-loop thermal effective potential.  The
important point, as can be seen in the high-temperature expansion, is
that, due to logarithmic terms associated with the fermionic sector of
the model, the effective Higgs quartic coupling can become negative at
a certain temperature, thus creating a ``runaway" behavior that is
stabilized by the higher-dimension operators.~\footnote{In some
examples the Higgs quartic at $T_c$ is positive but sufficiently small
to allow for the SM cubic term to induce the desired ``runaway"
behavior.} This realizes the basic idea explained in the introduction.
It also makes it manifest that the details of the phase transition are
UV dependent, since the higher-dimension operators in the Higgs
potential are crucial and depend on the undetermined $\bar{\gamma}$
and $\bar{\delta}$.  In order to show that a strong phase transition
can indeed be realized, we turn to a simple UV model that serves as a
``proof of existence."

The model we will focus on introduces the following new fields (using
the notation of Ref.~\cite{ABDF}, but see also~\cite{VFModel})
\beq
(\psi,\psi^c)
\sim (1,2)_{\pm \frac{1}{2}}~, \quad
(\chi,\chi^c)
\sim (1,1)_{\mp 1}~,
\label{psichi}
\eeq
where the charges correspond to the SM $SU(3) \times SU(2)\times
U(1)_Y$ quantum numbers.  These charges allow the following mass terms
for the new fermions
\beq
- {\cal L}_m = m_\psi \psi \psi^c + m_\chi \chi \chi^c + y H \psi \chi +
y_c H^\dagger \psi^c \chi^c + {\rm h.c.}
\label{Lmass}
\eeq
In the following we will assume, for simplicity, that $y = y_c$, in
which case we have two mass eigenstates with electric charges $|Q_i| =
1$ and squared masses given by
\bea
\hspace{-4mm}
m^2_{1,2}(\phi) &=& 
\frac{1}{2} \left(m_\psi^2 + m_\chi^2 \right) + \frac{1}{2} y^2 \phi^2
\label{m12}
\\
&\mp& \frac{1}{2}(m_\psi + m_\chi) \sqrt{(m_\psi - m_\chi)^2 + 2 y^2 \phi^2}~.
\nonumber
\eea
The spectrum also contains a neutral state $N$ with mass $m_N=m_\psi$.

We will consider the above model in the limit where $m_\psi \gg y v$
(while $m_\chi$ is at the EW scale).  In this limit, it is appropriate
to integrate out the heavy state with mass of order $m_2(\phi) \approx
m_\psi$, which can then be seen to generate the operator in
Eq.~(\ref{dim5}) with
\beq
G_m =  \frac{Z_m y^2}{2(m_\psi - m_\chi)}~.
\label{Gm}
\eeq
At tree-level one finds $Z_m = 1$ and one can check that
Eq.~(\ref{dim5}) reproduces $m_1(\phi)$ in Eq.~(\ref{m12}) to order
$\phi^2$.  We allow for a non-trivial factor $Z_m$ to investigate the
possible impact of higher-order loop corrections at the matching
scale, since we will later be interested in a region where the Yukawa
coupling, $y$, is sizeable.  The model also predicts that
$\bar{\gamma} = \bar{\gamma}_{\rm th} + \bar{\gamma}_{\rm RG}$ and
$\bar{\delta} = \bar{\delta}_{\rm th} + \bar{\delta}_{\rm RG}$ (see
the appendix), where
\bea
\bar{\gamma}_{\rm th} &=& \frac{Z_\gamma y^6}{16 \pi^2 } \, \frac{m_\psi (m_\psi^2 + 7 m_\chi m_\psi - 2 m_\chi^2)}{(m_\psi - m_\chi)^5}~,
\label{gammath} \\
\bar{\delta}_{\rm th} &=& - \frac{Z_\delta y^8}{48 \pi^2} \, \frac{7 m_\psi^3 + 27 m_\chi m_\psi^2 -4 m_\chi^3}{(m_\psi - m_\chi)^7}~,
\label{deltath}
\eea
are the threshold contributions induced at $\mu = m_\psi$, and
\bea
\bar{\gamma}_{\rm RG} &\approx& - \frac{3 G_m^3 m_\chi}{2\pi^2} \ln \left( \frac{m_\psi^2}{\mu^2} \right)~,
\label{gammaRG} \\
\bar{\delta}_{\rm RG} &\approx& \frac{G_m^4}{2 \pi^2} \ln \left( \frac{m_\psi^2}{\mu^2} \right)~,
\label{deltaRG}
\eea
are the running contributions between $m_\psi$ and a lower scale
$\mu$.  Here, we parametrized possible higher-order loop effects at
the matching scale through the factors $Z_\gamma$ and $Z_\delta$ (at
lowest order, $Z_\gamma = Z_\delta = 1$).

\begin{figure}
\includegraphics[width=0.4\textwidth]{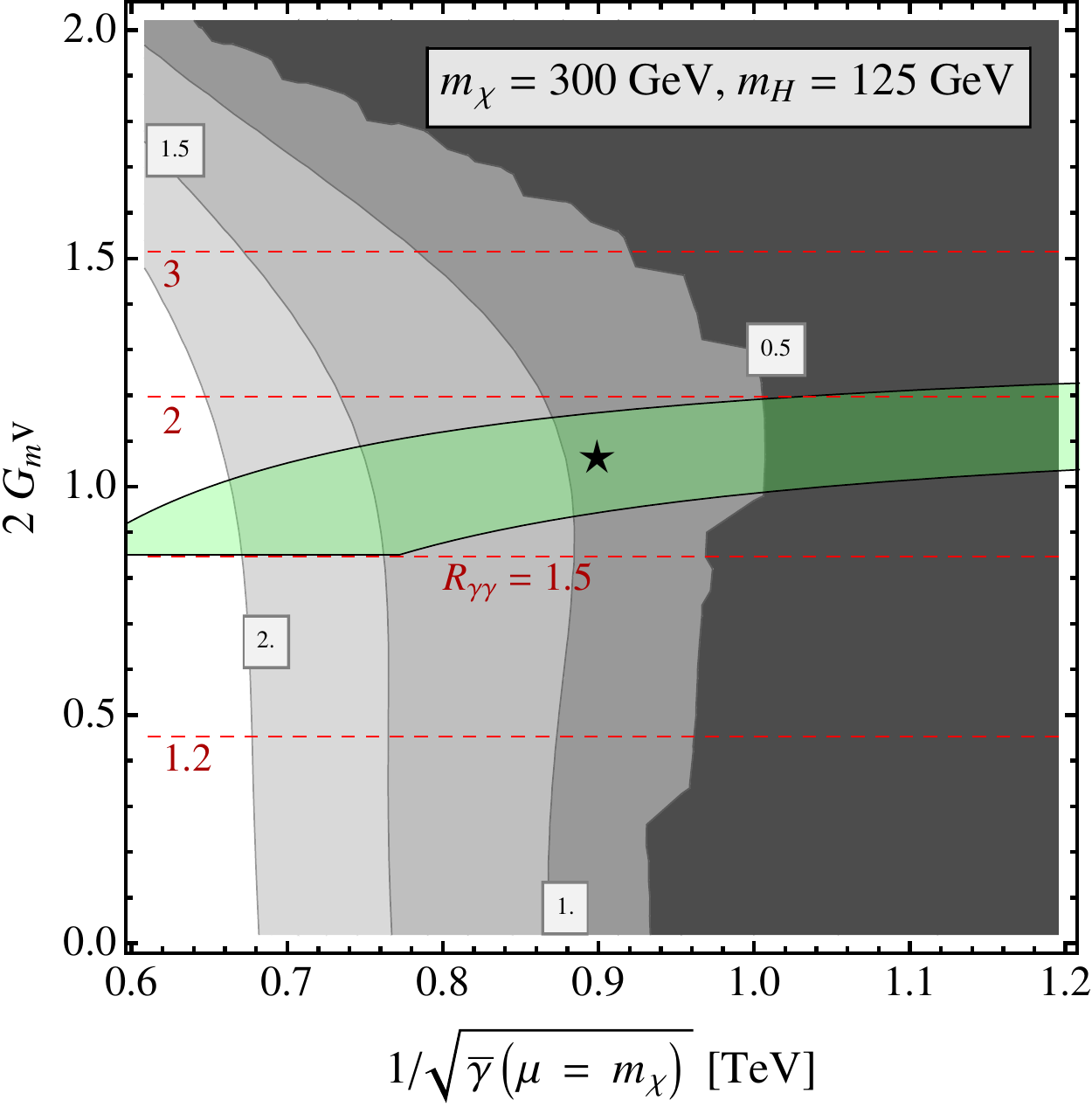}
\caption{Contours of constant $\phi_c / T_c$ in the EFT (gray) defined
by Eqs.~(\ref{V0})-(\ref{F1}), together with the diphoton enhancement
(dashed red horizontal lines).  The star corresponds to the benchmark
point in the UV model, while the light green region (narrow horizontal
band) corresponds to a 20\% variation in $Z_m$ and $Z_\gamma$ in that
benchmark.  Up to a sign, the vertical axis is the effective Yukawa
coupling of the light fermion to the Higgs (see text).}
\label{fig1}
\end{figure}

In Fig.~\ref{fig1}, we show the contours of $\phi_c / T_c$ in the
plane of $G_m v$ versus (the low-energy) $\sqrt{1/\bar{\gamma}}$, in a
model-independent analysis based on Eqs.~(\ref{V0})-(\ref{F1}), fixing
$m_\chi = 300~{\rm GeV}$.  We also indicate with a star a benchmark
point corresponding to the UV model discussed above [i.e.~using
Eqs.~(\ref{Gm})-(\ref{deltaRG})].  This benchmark has $m_\chi =
300~{\rm GeV}$, $m_\psi = 4~{\rm TeV}$ and $y = 4$.  For $Z_m =
Z_\gamma = Z_\delta = 1$, one has $m_1(v) \approx 170~{\rm GeV}$
(satisfying a naive LEP 2 bound of $\sim 100~{\rm GeV}$), and the
phase transition occurs at a critical temperature $T_c \approx
150~{\rm GeV}$ when $\phi_c \approx 140~{\rm GeV}$ so that $\phi_c /
T_c \approx 0.93$.\footnote{We note that the same qualitative features
can be obtained by a naive application of the Coleman-Weinberg
potential to the full theory defined by Eqs.~(\ref{psichi}) and
(\ref{Lmass}) [for the benchmark point one finds $\phi_c \approx
128~{\rm GeV}$ and $T_c \approx 146~{\rm GeV}$, so that $\phi_c / T_c
\approx 0.88$].  However, the EFT analysis clarifies the origin of the
strongly first-order EWPhT, allows us to understand the large
logarithms, and permits a simple estimate of the higher-order loop
effects at the matching scale.} When the Higgs contributions are
included, we find $\phi_c / T_c \approx 1.1$.  Note also that at $\phi
= \phi_c$ the light fermion has a mass $m_1(\phi_c) \approx 260~{\rm
GeV}$ and is heavier than at $T=0$, which is an effect opposite to the
case considered in Ref.~\cite{CMQW}.

 The first-order phase transition arises as described in the
 introduction, relying on a negative quartic coupling at finite
 temperature, stabilized by the dimension-6 operator in
 Eq.~(\ref{V0}).  The mechanism depends mainly on the fermion
 contributions (both $\chi$ and the top quark).  In particular, even
 if the contributions from the $W$ and $Z$ are ignored we still find a
 strongly first-order phase transition.  Thus, unlike other scenarios,
 a term cubic in $\phi$ need not play a crucial role.\footnote{We do
 not include the ``daisy resummation"~\cite{Dolan:1973qd} which would
 somewhat affect the cubic terms from the SM, but we expect its impact
 to be relatively minor.} We note that at very small $G_m v$, one
 recovers the scenario discussed in~\cite{Grojean:2004xa}, as seen
 from Eq.~(\ref{beta}) when the light fermion contributions are
 decoupled and the negative contribution to the quartic due to $\beta$
 is dominated by the (positive) $\bar{\gamma}$ term.  We also note
 that for values of $\bar{\gamma}$ larger than exhibited in the
 figure, a minimum at the origin may develop even at $T=0$, in which
 case one should make sure that the EWSB minimum is the global
 one~\cite{Grojean:2004xa}.  In this region, however, the EFT may not
 give an accurate description of the physics, so we explicitly exclude
 it.

We have also checked in selected examples the efficiency with which
the nucleation process takes place by computing numerically the bounce
action, $S_E(T)$, and checking that $B(T_n) \equiv S_E(T_n) / T_n$ can
reach the desired range $130 \lesssim B(T_n) \lesssim 140$, for
nucleation temperatures $T_n$ around $100~{\rm GeV}$ even when the
phase transition is sufficiently strong to allow for EWBG. When
$\phi_c / T_c$ becomes too large, however, the nucleation rate becomes
exponentially suppressed.  This will limit, but not exclude,
phenomenologically interesting regions with a strongly first-order
phase transition.

As illustrated by the benchmark example, within the UV model, a
sizeable underlying Yukawa coupling is required.\footnote{We also
found regions of parameter space, within the UV model of \eq{Lmass},
with $y\sim 2.5$ and roughly degenerate $m_\chi$ and $m_\psi$ of order
a few hundred GeV (with the same sign), where one could achieve a
strongly first-order phase transition for $m_1\gsim 100$~GeV. However,
such regions of parameter space give rise to suppression of the Higgs
diphoton rate, which is disfavored by the current data from the LHC.}
Hence, we also show in the figure a contour around the previous
benchmark point exhibiting the result of varying $Z_m$ and $Z_\gamma$
within 20\% to provide a feel for the sensitivity to the higher-order
loop effects at the matching scale $m_\psi$.  We see that there is a
significant sensitivity of $\phi_c/T_c$, especially to $Z_\gamma$
which is responsible for the horizontal variation in the plot, as it
affects the coefficient of the stabilizing dimension-6 operator.

We also note that after EWSB the operator in Eq.~(\ref{dim5}) induces
an effective Yukawa coupling between the new fermion and the Higgs
boson given by $y_{\rm eff} = -2G_m v$.  The requirement of a strongly
first-order phase transition implies that this Yukawa coupling be of
order one.  Thus, we can expect a relevant modification of the Higgs
branching fractions, in particular the $H \to \gamma\gamma$ rate when
$\chi$ is electrically charged.  Interestingly, the fact that $G_m$ is
positive means that the new fermion mass decreases as the Higgs vev
increases, which implies an enhancement of the diphoton rate
\cite{Carena:2012xa}.  In the presence of a charged vector-like
fermion, the ratio of the branching fraction into the diphoton final
state to its SM value is given by
\beq
R_{\diphot} \simeq \left|1 - \frac{F_{1/2}(\tau_1) \, Q_1^2}{F_{\rm SM}}
\frac{\partial\ln m_1(v)}{\partial \ln v}
\right|^2~,
\label{R2gam}
\eeq
where $F_{\rm SM}\simeq -6.49$ is associated with the SM amplitude,
$\tau_1 = 4m^2_1/m^2_H$, and $F_{1/2}(\tau)$ is the familiar loop
function (see e.g.~Ref.~\cite{Hunter}).  In Fig.~\ref{fig1}, we have
also superimposed the lines of constant $R_{\diphot}$ (dashed-red
lines).  As one can see, the region with a strongly first-order phase
transition corresponds to values of $R_{\diphot}>1$ in the vicinity of
the benchmark point.

Also, the dimension-6 operator of Eq.~(\ref{V0}) affects the Higgs
boson triple coupling $V'''(v) = 3m^2_H/v+8\bar{\gamma}v^3$, where
\eq{V0} is used, ignoring the $\phi^8$ operator; the first term is the
usual SM contribution.  As discussed previously and shown in
Fig.~\ref{fig1}, for a first order EWPhT a positive $\bar\gamma \sim
(1~{\rm TeV})^{-2}$ is required.  This corresponds to increasing the
SM prediction of the triple Higgs coupling by $\ord{1}$.  Naively,
this increase in the Higgs triple coupling would indicate an increased
signal rate for $gg\rightarrow HH$ at the LHC. However, Higgs pair
production proceeds through destructively interfering top quark box
and off-shell $s$-channel Higgs boson amplitudes, with the box diagram
being the dominant contribution.  Hence, a moderate increase in the
Higgs triple coupling increases the $s$-channel Higgs contribution and
paradoxically decreases the LHC Higgs pair production rate.  For
$(600~{\rm GeV})^{-2}\gtrsim\bar{\gamma}\gtrsim(900~{\rm GeV})^{-2}$,
the pair production rate is expected to be $40-60\%$ of the SM
prediction at the LHC. With full SM strength, it is estimated to take
around several ab$^{-1}$ of data at the 14 TeV LHC to exclude a
$V'''(v)=0$ at $90\%$ CL~\cite{Baur:2003gp}, making this measurement
very challenging at the LHC. However, with 2 ab$^{-1}$ of data, a
$500$ GeV and $1$ TeV ILC are expected to measure the Higgs
self-coupling with an accuracy of $\sim44\%$ and $\sim17\%$
respectively~\cite{ILC}.  Hence, with a sufficient amount of data, the
ILC can examine one of the key implications of this EWPhT scenario.
In addition, a measurement of the light fermion mass $m_1$ and the
diphoton rate could be used to infer $m_\chi$ and $G_m v$ within the
framework.  This, together with some knowledge about $\bar{\gamma}$
from the Higgs boson triple coupling could identify the relevant
region in Fig.~\ref{fig1}, and start giving information regarding the
nature of the phase transition and its strength.

The model in \eq{psichi} does not allow for the decay of the light
charged fermion.  To avoid this situation (which could lead to severe
bounds on $m_1$), we could either assume that ($\psi, \chi$) mix with
the SM leptons or else postulate another heavy vector-like lepton $n$
with no SM charges \cite{ABDF}.  In the latter case, additional terms
of the form $H\psi n$ result in the appearance of two neutral states,
$n_{1,2}$, where $n_1$ could be lighter than $\chi$ and lead to
$\chi\to n_1 W$ ($W$ on- or off-shell) .  The new neutral states can
enhance the strength of the EWPhT, without affecting $R_{\diphot}$.

A second comment refers to a potential instability of the $T=0$ EFT
potential at large field values.  Using the SM RG equations together
with the contributions to the $\beta$-function discussed in the
appendix, and using for illustration the point marked by a star in the
figure, one can check that the Higgs quartic coupling becomes negative
at a scale of about $700~{\rm GeV}$ .  However, this regime is outside
the range of validity of the EFT analysis, and higher powers of $\phi$
can play a crucial role.  In fact, a naive application of the
Coleman-Weinberg potential to the full UV model suggests that when the
infinite tower of operators is resummed the instability is pushed to
about $3.5~{\rm TeV}$.  Although new (bosonic) degrees of freedom
would likely be required at this scale, they need not affect the
physics of the EWPhT or the diphoton rate.

Finally, we would like to comment on the connection to previous works
that share some elements of our setup and discussion.
Ref.~\cite{Grojean:2004xa} has examined the effects on the EWPhT
caused by higher dimension operators in the Higgs potential (see also
Refs.~\cite{Bodeker:2004ws,Delaunay:2007wb} for further work in this
direction).  In particular, they considered how a
\textit{zero-temperature} negative quartic could be stabilized by a
dimension-6 operator in their setup.  While this is a qualitative
feature also observed in our mechanism, we point out that in our model
the negative quartic is induced by the new fermion (with some help
from the top quark) at high temperatures and is not a feature of the
$T=0$ potential.  In fact, the analysis of Ref.~\cite{Grojean:2004xa}
ignored thermally induced quartic terms, although they were fully
included in Ref.~\cite{Delaunay:2007wb}.  Without the new fermionic
contribution, the cubic term from the SM would still play an essential
role.

As mentioned in the introduction, Ref.~\cite{CMQW} has also studied
how new heavy fermions with large couplings to the Higgs can lead to
an enhanced first order phase transition.  The mechanism studied in
Ref.~\cite{CMQW} is mainly based on transfer of entropy as the new
fermions get {\it heavier} and decouple from the plasma after EWSB. We
note that our results do not rely on this thermodynamic effect since
the relevant weak scale fermion in our mechanism gets {\it lighter}
and remains active in the thermal bath after EWSB. Note also that this
feature, i.e.~the decrease in the fermion mass as $\phi$ increases, is
the reason that our mechanism also leads to an enhancement in the
diphoton branching fraction of the Higgs, while the fermion couplings
in Ref.~\cite{CMQW} would lead to a suppression.

To conclude, we have shown that the tendency of fermions, that couple
with $\ord{1}$ strength to the Higgs, to drive the Higgs quartic
coupling negative can be related to a strongly first-order EWPhT. We
provided a concrete realization where the Higgs potential is
stabilized by higher-dimension operators that arise from a system of
heavy and light vector-like fermion pairs.  We note that the main
ingredients (one fermion with mass in the multi-TeV scale, a second
fermion parametrically lighter, and a semi-perturbative underlying
Yukawa interaction) can naturally arise in scenarios based on a warped
extra dimension.  This can potentially establish a connection between
our observation and the physics of EWSB, as well as the solution to
the hierarchy problem.  Furthermore, the same fermions can also
enhance the rate for $H\to \diphot$, as may be suggested by the early
LHC data.  We find that the Higgs diphoton decay rate and the strength
of the EWPhT can be correlated, as shown in the simple model above.
Also, a typical consequence of our setup is an enhancement of the
triple Higgs boson coupling that could be probed at the LHC or, more
likely, at a future lepton collider.

\vspace*{3mm}

\appendix*

In this appendix, we derive the effective theory at 1-loop order,
valid below the heavy fermion mass $m_2 \approx m_\psi$ in the model
defined by Eqs.~(\ref{psichi}) and (\ref{Lmass}).  We work in the
$\overline{{\rm MS}}$ scheme, matching the $\phi$ correlators up to
8th order.  In the UV theory, which describes both fermionic states,
$\psi$ and $\chi$, the correlators can be read from the
Coleman-Weinberg potential in the $\overline{{\rm MS}}$ scheme
\bea
V_{\rm UV} &=& V_{\rm tree} - \frac{4}{64\pi^2} \sum_{i=1,2} m_i^4(\phi) \left[ \ln \left( \frac{m^2_i(\phi)}{\mu^2} \right) - \frac{3}{2} \right],
\nonumber
\eea
where the $m_i$ are given in Eq.~(\ref{m12}).  Similarly, the
correlators in the effective theory are read from
\bea
\hspace{-5mm}
V_{\rm EFT} &=& V_{0} - \frac{4 m_1^4(\phi)}{64\pi^2} \left[ \ln \left( \frac{m^2_1(\phi)}{\mu^2} \right) - \frac{3}{2} \right],
\label{VMSEFT}
\eea
where $V_0$ is defined in Eq.~(\ref{V0}), and we include only the
light state, with a mass given in Eq.~(\ref{m12}).  Comparing the
$\phi^6$ and $\phi^8$ terms in both theories, and requiring that they
agree at $\mu = m_\psi$, fixes the corresponding threshold corrections
as given in Eq.~(\ref{gammath}) and (\ref{deltath}).  The matching
contributions to the mass parameter and Higgs quartic are not
interesting since they will be traded for $v$ and $m_H$.  However, the
running of all the parameters below $m_\psi$ is of interest.  The
$\beta$-functions can be derived from the requirement that the
effective potential in Eq.~(\ref{VMSEFT}) be RG invariant:
\bea
\sum_a \left( \beta_a \frac{\partial}{\partial \lambda_a} - \gamma_\phi \phi \frac{\partial}{\partial \phi} \right) V_0 = 
- \frac{4}{32\pi^2} m^4_1(\phi)~,
\nonumber
\eea
where $\gamma_\phi$ is the $\phi$ anomalous dimension, and we may use
$m_1(\phi) = m_\chi - G_m \phi^2$, which corresponds to keeping only
the operator (\ref{dim5}) in the EFT. Noting that this operator does
not induce any wavefunction renormalization for $\phi$ at one loop,
this leads to the $\beta$-functions in the $\overline{{\rm MS}}$
scheme:
\beq
\Delta \beta_\lambda = - \frac{3G_m^2 m_\chi^2}{\pi^2}~, 
\hspace{5mm}
\beta_\gamma = \frac{3G_m^3 m_\chi}{\pi^2}~, 
\hspace{5mm}
\beta_\delta = - \frac{G_m^4}{\pi^2}~.
\nonumber
\eeq
The operator in Eq.~(\ref{dim5}) renormalizes both $m_\psi$ and $G_m$
itself.  However, the first effect vanishes in the EW symmetry
preserving limit (when the Higgs is massless), while the second effect
is of second order in the small quantity $G_m$, thus leading to a
rather mild $\mu$-dependence.  Neglecting these effects, as well as
the renormalization due to the weak interactions, the RG equations can
be solved immediately to produce Eqs.~(\ref{gammaRG}) and
(\ref{deltaRG}).

We also provide the effective potential in the EFT using the
high-temperature expansion and neglecting the weak gauge boson
contributions.  Note that the heavy fermion state will be efficiently
Boltzmann suppressed at temperatures relevant to EWPhT and hence it
can be ignored.  Up to a constant term,
\bea
V(\phi) = \frac{1}{2} \mu^2_{\rm eff} \phi^2 + \frac{1}{4} \lambda_{\rm eff} \phi^4 + \frac{1}{6} \gamma_{\rm eff} \phi^6 + \frac{1}{8} \delta_{\rm eff} \phi^8 + {\cal O}(\phi^{10})~,
\nonumber
\eea
where
\bea
\mu^2_{\rm eff} &=& - \frac{m_H^2}{2} + \frac{T^2}{12} \left( 3 y^2_t - 4 G_m m_\chi \right)
- \frac{3 y^4_t v^2}{16 \pi^2} 
\nonumber \\
&+& \alpha + \frac{G_m m_\chi^3}{2\pi^2} \ln \left( \frac{A_F T^2}{\mu^2} \right)~,
\nonumber \\
\lambda_{\rm eff} &=& \frac{m_H^2}{2v^2} + \beta - \frac{3y^4_t}{16\pi^2} \ln\left( \frac{2A_F T^2}{y^2_t v^2} \right)
\nonumber \\
&+& \frac{1}{3} G_m^2 T^2 - \frac{3G_m^2 m_\chi^2}{2\pi^2} \ln \left( \frac{A_F T^2}{\mu^2} \right)~,
\nonumber \\
\gamma_{\rm eff} &=& \bar{\gamma} + \frac{3G_m^3 m_\chi}{2\pi^2} \ln \left( \frac{A_F T^2}{\mu^2} \right)~,
\nonumber \\
\delta_{\rm eff} &=& \bar{\delta} - \frac{G_m^4}{2\pi^2} \ln \left( \frac{A_F T^2}{\mu^2} \right)~.
\nonumber
\eea
$\alpha = \alpha(m^2_1(v))$ and $\beta = \beta(m^2_1(v))$ are defined
in Eqs.~(\ref{alpha}) and (\ref{beta}) [by definition, these include
only the new light fermion field], $\bar{\gamma}$ and $\bar{\delta}$
are defined by Eq.~(\ref{V0}), and $A_F = \pi^2 e^{-2\gamma_E}$ with
$\gamma_E$ the Euler-Mascheroni constant.  We emphasize that the above
expression is independent of the UV completion.  One can check that
the explicit $\mu$-dependence cancels out when Eqs.~(\ref{gammaRG})
and (\ref{deltaRG}) are used.  Note also that the contribution to the
thermal mass from the light fermion state is negative.  However, even
in the absence of the top quark, at sufficiently high temperatures,
the heavy fermion (or more generally the heavy states in the UV
completion) gives a positive contribution to the thermal mass that
restores the EW symmetry, as expected.  Thus, the above potential has
a region of validity in temperature with both upper and lower bounds.
For temperatures of order the EW scale it should provide an
appropriate description of EWPhT.

\vspace*{2mm}

\acknowledgments


We thank D.~Marzocca, M.~Serone and A.~Urbano for pointing out an
error in a previous version of this work.  This led us to consider a
different region of parameter space, where the original conclusions
could be obtained.  We also thank Mariano Quir\'{o}s for comments on
the revised manuscript, and Sally Dawson for helpful conversations.  
This work is supported by the US Department
of Energy under Grant Contract DE-AC02-98CH10886.



\begin{thebibliography}{99}

\bibitem{ATLASHiggs}
 G.~Aad {\it et al.}  [ATLAS Collaboration],
  Phys.\ Lett.\ B {\bf 716}, 1 (2012)
  [arXiv:1207.7214 [hep-ex]].

\bibitem{CMSHiggs}
 S.~Chatrchyan {\it et al.}  [CMS Collaboration],
  Phys.\ Lett.\ B {\bf 716}, 30 (2012)
  [arXiv:1207.7235 [hep-ex]].

\bibitem{VF}
  S.~Dawson and E.~Furlan,
  Phys.\ Rev.\ D {\bf 86}, 015021 (2012)
  [arXiv:1205.4733 [hep-ph]];
%
  N.~Bonne and G.~Moreau,
  arXiv:1206.3360 [hep-ph];
%
  H.~An, T.~Liu and L.~-T.~Wang,
  arXiv:1207.2473 [hep-ph].
%
  A.~Joglekar, P.~Schwaller and C.~E.~M.~Wagner,
  arXiv:1207.4235 [hep-ph].
%
  H.~Davoudiasl, H.~-S.~Lee and W.~J.~Marciano,
  arXiv:1208.2973 [hep-ph];
%

\bibitem{Carena:2012xa} 
  M.~Carena, I.~Low and C.~E.~M.~Wagner,
  JHEP {\bf 1208}, 060 (2012)
  [arXiv:1206.1082 [hep-ph]].

\bibitem{VFModel}
  L.~G.~Almeida, E.~Bertuzzo, P.~A.~N.~Machado and R.~Z.~Funchal,
  arXiv:1207.5254 [hep-ph];
%
  J.~Kearney, A.~Pierce and N.~Weiner,
  arXiv:1207.7062 [hep-ph];
%
  M.~B.~Voloshin,
  arXiv:1208.4303 [hep-ph];

\bibitem{ABDF}
  N.~Arkani-Hamed, K.~Blum, R.~T.~D'Agnolo and J.~Fan,
  arXiv:1207.4482 [hep-ph].

\bibitem{CMQW}
  M.~S.~Carena, A.~Megevand, M.~Quir\'os and C.~E.~M.~Wagner,
  Nucl.\ Phys.\ B {\bf 716}, 319 (2005)
  [hep-ph/0410352];

\bibitem{Espinosa:1993yi} 
  J.~R.~Espinosa, M.~Quir\'os and F.~Zwirner,
  Phys.\ Lett.\ B {\bf 307}, 106 (1993)
  [hep-ph/9303317].

\bibitem{Carena:1996wj} 
  M.~S.~Carena, M.~Quir\'os and C.~E.~M.~Wagner,
  Phys.\ Lett.\ B {\bf 380}, 81 (1996)
  [hep-ph/9603420].

\bibitem{EWPTDP}
For recent work that explores the connection between the EWPhT and the $H \to \gamma \gamma$ rate, invoking extra bosonic degrees of freedom, see
  D.~J.~H.~Chung, A.~J.~Long and L.~-T.~Wang,
  arXiv:1209.1819 [hep-ph], and 
  W.~Huang, J.~Shu and Y.~Zhang,
  arXiv:1210.0906 [hep-ph].

\bibitem{Grojean:2004xa} 
  C.~Grojean, G.~Servant and J.~D.~Wells,
  Phys.\ Rev.\ D {\bf 71}, 036001 (2005)
  [hep-ph/0407019].

\bibitem{Delaunay:2007wb} 
  C.~Delaunay, C.~Grojean and J.~D.~Wells,
  JHEP {\bf 0804}, 029 (2008)
  [arXiv:0711.2511 [hep-ph]].

\bibitem{Anderson:1991zb} 
  G.~W.~Anderson and L.~J.~Hall,
  Phys.\ Rev.\ D {\bf 45}, 2685 (1992).

\bibitem{Quiros:1999jp}
  M.~Quir\'os,
  hep-ph/9901312.

\bibitem{Cahill:1993mg} 
  K.~E.~Cahill,
  Phys.\ Rev.\ D {\bf 52}, 4704 (1995)
  [hep-ph/9301294].

\bibitem{Dolan:1973qd} 
  L.~Dolan and R.~Jackiw,
  Phys.\ Rev.\ D {\bf 9}, 3320 (1974).

\bibitem{Hunter} 
 J.~F.~Gunion, H.~E.~Haber, G.~Kane and S.~Dawson,
 \textit{The Higgs Hunter's Guide,}
 Addison Wesley Publishing Company (1990).


\bibitem{Baur:2003gp} 
  U.~Baur, T.~Plehn and D.~L.~Rainwater,
  Phys.\ Rev.\ D {\bf 69}, 053004 (2004)
  [hep-ph/0310056];
  K.~Jakobs,
  Eur.\ Phys.\ J.\ C {\bf 59}, 463 (2009).

\bibitem{ILC} 
 H.~Baer~et~al., \textit{ILC DBD,} Physics at the International Linear
 Collider, to be published in the ILC Detailed Baseline Design Report.

\bibitem{Bodeker:2004ws} 
  D.~Bodeker, L.~Fromme, S.~J.~Huber and M.~Seniuch,
  JHEP {\bf 0502}, 026 (2005)
  [hep-ph/0412366].


\end{thebibliography}
\end{document}